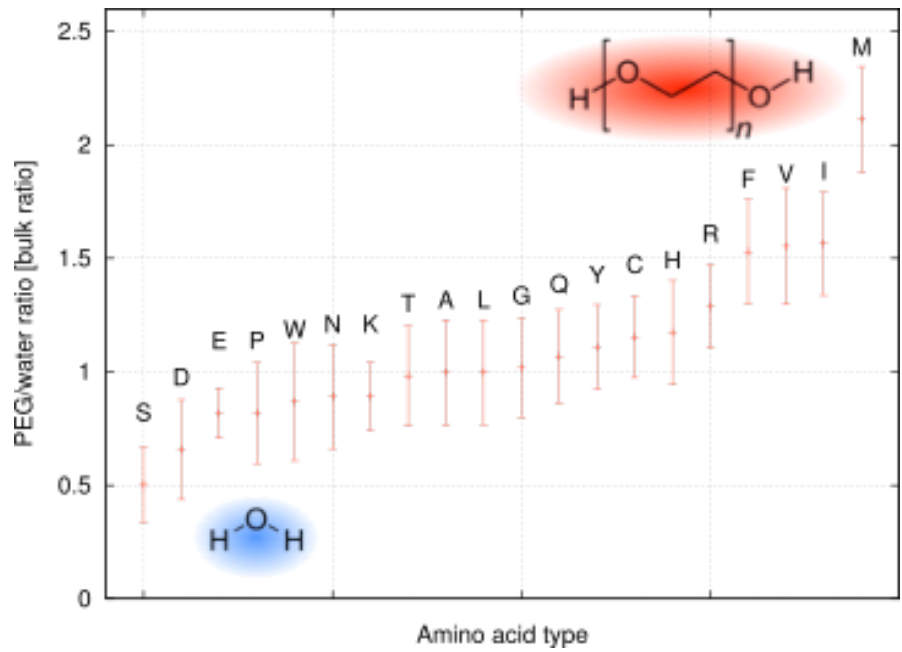

**Table of content entry**

The affinity of protein surface amino acids for poly(ethylene glycol)(PEG)(reported by the PEG/water ratio in their vicinity) is used to build a quantitative model of protein adsorption on PEGylated nanoparticles.

# Journal Name

## COMMUNICATION

# Protein corona composition of PEGylated nanoparticles correlates strongly with amino acid composition of protein surface

Giovanni Settanni*[1,2], Jiajia Zhou[1], Tongchuan Suo[1], Susanne Schöttler[3,4], Katharina Landfester[3], Friederike Schmid*[1], Volker Mailänder*[3,4]

**Extensive molecular dynamics simulations reveal that the interactions between proteins and poly(ethylene glycol)(PEG) can be described in terms of the surface composition of the proteins. PEG molecules accumulate around non-polar residues while avoiding polar ones. A solvent-accessible-surface-area model of protein adsorption on PEGylated nanoparticles accurately fits a large set of data on the composition of the protein corona recently obtained by label-free proteomic mass spectrometry.**

Nanoparticles are being intensively investigated as possible drug carriers due to their ability to load drugs, and potentially deliver them selectively to the target (e.g. the cancer cells or tissues), opening the way to highly effective therapies with reduced side effects[1]. As soon as the nanoparticles come into contact with the blood stream they are covered by plasma proteins, forming the "protein corona". Numerous studies have demonstrated that the nature and composition of the protein corona determines the most important characteristics of the nanoparticles, for example, their ability to be taken up by cells or to stimulate an immune response, or their toxicity[2,3]. The composition of the protein corona can be measured experimentally, for example by liquid chromatography mass spectrometry[4,5]. However many other features, like the orientation of the adsorbed proteins and the degree of exposure of the proteins' functional epitopes are more difficult to access[6]. These features, as well as the determinants of the protein corona composition, are particularly important as they help to understand how the corona will interact with the host organism.

Poly(ethylene glycol) (PEG) has a wide variety of applications, which often involve its capacity to limit protein adsorption. PEG, immobilized to surfaces, greatly retards protein adsorption[7] and shows antifouling activity[8]. PEGylation of drugs and nanocarriers leads to an increment of their circulation half-lives, not only by increasing their hydrodynamic radius or thermal stability, but also by decreasing their susceptibility to phagocytosis[9]. These properties of PEG – often subsumed under the name of "stealth" effect – are essential for the effective targeting of nanocarriers[10]. The stealth effect has been generally explained by the large hydrophilicity of PEG which leads to the formation of a thick hydration layer that hinders interactions with the surrounding proteins. Notwithstanding its "stealth" properties, PEG is not totally inert. Indeed, several proteins have been detected in protein coronas of PEGylated nanocarriers after incubation in serum or plasma[11]. Recent data from label-free proteomic mass spectrometry of PEG-coated nanoparticles[5], show that their coronas have a very specific protein composition, which is significantly different from the one observed in plasma. These data suggest that PEG may not limit uniformly the adsorption of proteins, but it may do so in a protein-dependent way, with few proteins showing a highly preferential binding affinity for PEG-coated nanoparticles. The possible reasons of the observed preferential binding are not well understood, yet.

Molecular dynamics (MD) simulations have been extensively used to describe protein/peptide adsorption on several different surfaces as recently reviewed in ref. [12] and [13]. Protein adsorption on surfaces coated with hydrophilic polymers like PEG provide a further challenge due to their brush structure. In this case, depending on their size, the proteins may partially or completely penetrate into the polymer brush formed on the coated surface[14]. Here, we evaluate the possible origin of the selective adsorption of proteins on the surface of PEG-coated nanoparticles and show how it can be related to and predicted by the amino acid composition of the protein surface. We address the problem by performing atomistic MD simulations of selected plasma proteins immersed in PEG-water mixtures characterized by different PEG lengths and concentrations. These simulations help us understand how PEG molecules distribute on the surface of the proteins and relate PEG distribution to protein surface properties. From the simulation results we infer the possible relationship between the concentration of proteins in the nanoparticle corona and the amino acid composition of their surface. Finally we verify the inferred relationship on a large dataset of proteins, using only available structural information without performing further simulations.

[1]Institute of Physics, Johannes Gutenberg University Mainz,
[2]Max Planck Graduate Center with the Johannes Gutenberg University Mainz,
[3]Max Planck Institute for Polymer Research, Mainz,
[4]University Medical Center, Dermatology Clinic, Johannes Gutenberg University Mainz
*e-mail: settanni@uni-mainz.de, friederike.schmid@uni-mainz.de, mailaend@mpip-mainz.mpg.de

Electronic Supplementary Information (ESI) available: Details of the methodology used for the simulations and the analysis of the data including two supplementary figures and two supplementary tables.

The simulations were prepared by immersing selected plasma proteins into periodic simulation boxes filled with pre-equilibrated PEG/water mixtures at physiological ionic strength (0.15M [NaCl]). The simulations were carried out using the program NAMD[15] and the charmm27 force field[16,17] including parameters for PEG[18]. Tip3p[19] was used as model for the explicit treatment of water. Further details about the simulations and the data analysis are provided in the supplementary information. In the course of the MD simulations, all the particles (water, PEG, ions and protein) can diffuse freely, thus allowing for the PEG molecules to probe many possible binding modes with the protein surface. This approach has been successfully used in the context of drug design to identify druggable cavities in proteins[20,21] and also to study protein-polymers interactions including PEG [22,23]. It is justified by the fact that PEG is not assumed to induce large conformational changes in the protein conformation[24]. We have purposely used short PEG chains (4 and 7 monomers), which diffuse fast, and allow for fast sampling of the possible binding modes. Given the very short persistence length of PEG (0.38nm) which is comparable with the distance between adjacent monomers, even the short PEG chains capture the flexibility of the longer PEG chains used in the experiments. Thus, in the polymer conformations bound to the protein surface the monomers are expected to occupy similar positions either if they belong to long or short PEG chains. If non-negligible interactions are formed between the protein (or parts of it) and PEG molecules, the concentration of PEG atoms in the vicinity of the protein (or the selected parts) will differ from the value in the bulk. Thus, we measure the ratio between the number of PEG atoms (excluding hydrogen atoms) and water molecules present within a certain distance (0.5 nm) from the protein (or the selected parts). A PEG/water ratio larger than the bulk values would indicate the presence of effective attractive interactions between PEG and protein (or the selected parts). We have analyzed 3 plasma proteins, human serum albumin (HSA), transferrin(TRF), and complement Cq1 subcomponent subunit C(CQ1C). These proteins were selected based on several criteria including the availability of crystallographic structure and either high (CQ1C) or low (HSA, TRF) experimentally observed concentration in the protein corona of PEG-coated nanoparticles[5]. We have also analyzed the possible effects of PEG concentration (from 0.04 to 0.12 g/ml) and length (4 or 7 monomers), as well as simulation box size. A list of all the simulated systems including number and length of runs is provided in table S1 in the supplementary information. The length of the simulations (200ns) is sufficient to reach statistical convergence of the PEG-protein interactions (see supplementary information). The simulations show that (i) the analyzed proteins have different affinity for PEG molecules, consistent with the experimental observation of the protein concentration in the nanoparticle corona; (ii) PEG molecules do not distribute uniformly on the protein surfaces. Instead, they tend to accumulate in specific regions (Figure 1a,b,c).

We assumed that these observations might reflect the heterogeneous amino acid composition of the protein surface in the analyzed proteins. Thus, we measured the residue specific affinity for PEG in all the simulations, i.e. the PEG/water ratio in the vicinity of each amino acid type, and normalized it to the bulk PEG/water ratio. It turns out that these values are surprisingly similar across the different proteins and simulation conditions, especially for the most solvent exposed residue types (Figure 1d) (for the less exposed residue types the statistical noise is large).

This finding led us to propose a simple model to describe PEG-protein interactions, based on few assumptions. The first assumption is that the total PEG/water ratio (in the vicinity) of a protein $P$ can be approximated as the sum of the contributions coming from each residue type weighted by the corresponding solvent-accessible surface area.

$$\text{PEGW}_{\text{tot}}(P) = \sum^{AA} \frac{\text{SASA}_{\text{AA}}(P)}{\text{SASA}_{\text{tot}}(P)} \cdot \text{PEGW}_{\text{AA}} \quad (1)$$
$$= \sum^{AA} \text{FSASA}_{\text{AA}}(P) \cdot \text{PEGW}_{\text{AA}}$$

where AA is the residue type (i.e., alanine, cysteine, aspartate, etc.), $\text{SASA}_{\text{AA}}(P)$ is the solvent-accessible surface area of all the residues of type AA in the protein $P$, $\text{SASA}_{\text{tot}}(P)$ is the total solvent-exposed surface area of the protein, $\text{FSASA}_{\text{AA}}(P)$ is the fraction of surface area exposed by residues type AA, and $\text{PEGW}_{\text{AA}}$ is the PEG/water ratio in the vicinity of the residue type AA (normalized to the bulk ratio), which, according to the analysis presented above, is approximately independent of the protein under investigation.

We tested this assumption by taking the $\text{PEGW}_{\text{AA}}$ averaged over different simulations (weighting the simulations according to the solvent-exposed-surface area of the amino acid type) and computing the approximated $\text{PEGW}_{\text{tot}}(P)$ with eq. (1) for each of the simulated proteins. We then compared this number with the actual $\text{PEGW}_{\text{tot}}(P)$ measured directly in the simulations. The comparison (Figure 1e) shows that the approximation reproduces within the error, the affinity of the protein for PEG (measured with the PEG/water ratio) for the three simulated proteins.

The second assumption that we make is that the average adhesion force (force per unit area) $F_A$ between a protein and a PEG-coated nanoparticle is a function of the PEG/water ratio.

$$F_A = F_A(\text{PEGW}_{\text{tot}}(P)) \quad (2)$$

In other words, a high PEG/water ratio of the protein would result in a strong adhesion force, due to the large amount of PEG molecules per unit area that are attracted in the vicinity of the protein. Although the core of the PEGylated nanoparticle may have an influence on the nature and quantity of adsorbed proteins[11], at this point we are neglecting it. The third assumption is that the concentration of proteins observed on the nanoparticle surface is proportional to the adhesion force and the concentration of the protein in solution (i.e., in plasma):

$$[P]_{\text{np}} \propto F_A(\text{PEGW}_{\text{tot}}(P)) \cdot [P]_{\text{plasma}} \quad (3)$$

where $[P]_{\text{np}}$ and $[P]_{\text{plasma}}$ are the concentrations of the protein $P$ on the nanoparticle and in plasma, respectively.

The second and third assumptions can be summarized by the following equation:

$$\log\left(\frac{[P]_{\text{np}}}{[P]_{\text{plasma}}}\right) = f(\text{PEGW}_{\text{tot}}(P)) \quad (4)$$

where the left hand side is a term which resembles a free energy of binding, and the right hand side is a generic function of the PEG/water ratio. The simplest choice for $f$, which we will adopt, is a linear function of the PEG/water ratio. Combining eq.(1) with eq. (4) and the linear dependence we obtain:

$$\log\left(\frac{[P]_{\text{np}}}{[P]_{\text{plasma}}}\right) \propto \sum^{AA} \text{FSASA}_{AA}(P) \cdot \text{PEGW}_{AA} \qquad (5)$$

In the following, we prove the validity of eq. (5) using available experimental data.

To test eq. (5) we obtained the protein concentrations $[P]_{\text{np}}$ and $[P]_{\text{plasma}}$ from proteomic mass spectrometry experiments[5], where the amount of different plasma proteins was measured both on PEG-coated nanoparticles and in plasma (see supplementary information for details). Then, if available, the structures of the proteins adsorbing on the nanoparticles were collected either from the Protein Data Bank (PDB) or from the homology model database[25]. This resulted in a database of 36 plasma proteins (Table S2 in supplementary information). Then, we computed the solvent exposed surface area of each residue type in each protein using VMD[26], which provided the $\text{FSASA}_{AA}(P)$ data in eq. (5).

Using in eq. (5) the $\text{PEGW}_{AA}$ determined with the simulations, we already observe a significant correlation coefficient (r=0.55, p<0.001, df=36)[27] between the left and right hand side of the equation (Figure 2a). On the other hand, the values obtained from the simulations may be biased by several factors including the short length of the PEG molecules, the approximate nature of the force field, the limited number of simulated proteins and the low statistical accuracy for the less exposed amino acids. For these reasons, starting from the average values obtained from the simulations, we optimized the $\text{PEGW}_{AA}$ to maximize the correlation coefficient between left- and right-hand side of eq. (5) using a bootstrap approach. Details about the fitting procedure are provided in the supplementary information. The resulting model shows a very high correlation coefficient (r=0.85) with the available data (Figure 2b). We proved the statistical significance of this correlation by generating a thousand artificial datasets, where the $\text{FSASA}_{AA}$ were randomized across the proteins, and applying the same bootstrap procedure to each set. Only 1.4% of the random datasets led to correlation coefficients higher than the one measured on the original data, proving the statistical significance of the observed correlation. This fact supports the validity of eq. (5).

As exposed more diffusely in the supplementary information, the bootstrap procedure used for fitting the experimental data involves, fitting the parameters to a subset of the proteins in the original data set (training set) and then using the fitted parameters also to predict the protein concentration of the remaining proteins (test set). This is repeated for many different random choices of the training and test set. Thus, the results presented here, which are averages over many choices of training and test sets show that the model is predictive and the high correlation coefficient provides an assessment for that.

The $\text{PEGW}_{AA}$ values obtained from the fit and reported in Table 1 (which do not deviate strongly from the average values obtained from the simulations, with correlation coefficient 0.69, see Figure S1 in supplementary information), provide a measure of the relative affinity of each residue type for PEG. They show, for example, that negatively charged residues tend to repel PEG, as expected by the presence of a negative partial charge concentrated on the oxygen atoms in PEG. Unexpectedly, however, also positively charged lysine, which generally provides the largest fraction of solvent-accessible surface to the analyzed proteins, shows the tendency to have lower density of PEG in its vicinity than in the bulk. A possible explanation of this phenomenon could be related to the large charge density of lysine. According to the CHARMM force field lysine's charge is concentrated on the zeta-nitrogen (NZ) and epsilon-carbon (CE) atoms (and bound hydrogen atoms), while, as a comparison, arginine's charge is spread over 5 heavy atoms (and bound hydrogen atoms). The large charge density may favor interactions with water over PEG, which has a smaller polarity.

On the other hand, the data show that PEG interacts in general more favorably with hydrophobic residues than with polar or hydrophilic residues. This is in agreement with previously published simulations of CI2 in PEG4 water mixtures[22] which showed the formation of a larger number of contacts between PEG and atoms with small partial charges compared to atoms with large partial charges. Similarly, sum frequency generation vibrational spectroscopy experiments[28] showed that the addition of free PEG in solution reduced the number of strong hydrogen bonds between lysozyme and water indicating the formation of interactions between PEG and the hydrophobic parts of the protein. However, it has been shown that free PEG and surface-bound PEG as found on nanoparticles may affect the protein hydration shell differently[28]. The discrepancy was attributed to the expected burial of the CH2 groups of PEG in the surface film but it may also be related to the interactions of the protein with the core of the nanoparticle. Indeed, it has been shown that, although providing qualitatively similar protein adsorption patterns, differences in the core of PEGylated nanoparticles may result in differences in the amount and type of the adsorbed proteins[11]. This may explain the partial discrepancy between the PEG/water ratio measured in the simulations and those obtained by fitting the experimental data.

The model proposed here is based on the assumption that the proteins in contact with the nanoparticle surface retain their structure, including the solvent exposed surface area of their amino acids. This assumption was verified experimentally for some of the proteins[29]. In addition, the simulated proteins did not show relevant conformational changes during the simulations with PEG. The C$\alpha$ root mean square deviation from the crystallographic structures averaged across the simulation runs remained below 0.2, 0.35 and 0.5 nm for C1QC, HSA and TRF, respectively, compatible with native state fluctuations for proteins of that size. However, if the proteins underwent unfolding upon adsorption, a better determinant for their adsorption properties could be their amino acid sequence composition. Then, eq. (5) could be modified by replacing the $\text{FSASA}_{AA}$ of the amino acids with their relative abundance in the sequence. We fitted this sequence-based model to the experimental data using the same bootstrap procedure used above and we obtained a correlation coefficient of 0.79 with the experimental data, which is lower than what was obtained using the solvent exposed surface area. This indicates that, on average, the fraction of solvent exposed surface area is a better predictor of protein adsorption on nanoparticles, and lets us conclude that the majority of the proteins examined here remain folded upon adsorption. However, we cannot exclude that some of the proteins may undergo unfolding upon adsorption.

The adsorbed-protein concentration data used in eq.(5) were obtained on $PEG_{44}$-coated polystyrene nanoparticles (PS-PEG44). These data are available also for $PEG_{110}$-coated polystyrene (PS-PEG110) nanoparticles[5]. PS-PEG44 and PS-PEG110 have on average 3500 and 2900 grafted chains per particle, a size of 117nm and 119nm, and a zeta-potential of 15mV and 8mV, respectively. While the total adsorbed proteins varied significantly in the two cases (0.77mg/m$^2$ and 1.24mg/m$^2$, respectively), the relative values of the left hand side of eq.5 ($\log([P]_{np}/[P]_{plasma})$) do not change significantly and show a correlation of 0.96 between the two nanoparticles' coronas over the set of 36 proteins we have analyzed. This, of course, does not include all the possible variations of PEGylated nanoparticles, however, it suggests that the quantity we have identified ($\log([P]_{np}/[P]_{plasma})$) is only weakly dependent on the characteristics of the PEGylated nanoparticle. In particular, the reported grafting density of PS-PEG44 and PES-PEG110[5] (3.0 and 3.8 nm$^2$ per PEG chain, respectively) indicates that PEG chains form a dense brush[30,31]. Protein adsorption pattern of PEGylated nanoparticles do not change significantly above a certain PEG density[11], thus the analysis presented here is possibly valid for PEGylated nanoparticles with similar or higher PEG densities than those analyzed here.

Although the main contribution (about 60%) to the PEG/water ratio in the vicinity of the protein comes from the most exposed amino-acids, we found that the $FSASA_{AA}(P)$ of few specific amino acid types correlate particularly strongly with the left hand side of eq. (5). One striking example is methionine. Indeed, the $FSASA_{MET}(P)$ alone shows a correlation of 0.62 with the left hand side of eq.(5), suggesting that proteins where methionine amino acids expose the largest fraction of surface area may more likely adsorb on PEG-coated nanoparticles. This also explains why methionine is the amino acid with the largest $PEGW_{AA}$ in both the fitted parameters and those derived from the simulations (see supplementary figure S1). The analysis of the trajectories shows an accumulation of PEG in the vicinity of exposed methionine side-chains (supplementary figure S2). Interestingly, clusterin, which is the protein that adsorbs most abundantly on the PEG-coated nanoparticles analyzed here, but is not part of our data set as it has not been structurally characterized yet, has a very high percentage of methionine amino acids in its sequence (3.5%) compared to the other proteins in the data set. A higher percentage of methionine is found only in fibrinogen beta chain (3.7%) and complement C1q subcomponent B subunit (3.6%), which also adsorb abundantly on the nanoparticle surface.

## Conclusions

The simulations presented here show that PEG-protein interactions can be described using a simple model based on the solvent-accessible surface area exposed by each amino acid type on the protein surface. The model has been applied to provide a simple and accurate description of the adsorption process of plasma proteins on the surface of PEGylated nanoparticles. The model, however, is not necessarily limited to plasma proteins and may be applicable to protein-PEG interactions occurring in other body fluids, as long as the main solvent is water. The model allows for making predictions on how PEG will interact with specific proteins, and is not necessarily limited to the corona of PEGylated nanoparticles. For example, it may allow for predicting which parts of the protein will be buried by PEG, or which binding epitopes will remain available for receptor binding. In turn, this may allow for a fast evaluation of the consequences of PEGylation in relation to the substrate-specific targeting needs. Although, very well-established as drug modifier, PEG has several shortcomings including interfering with cellular uptake and triggering immune response[32]. Alternatives to PEG are being investigated by several groups[5,32]. We expect our approach to be transferable to the description of other protein-nanocarrier interactions.

***Acknowledgments*** GS acknowledges support from the Max Planck Graduate Center (MPGC). We gratefully acknowledge support with computing time from HPC facility Mogon at the University of Mainz, and the High Performance Computing Center Stuttgart (ACID 44059). This work was supported by the DFG/SFB1066 ('Nanodimensionale polymere Therapeutika fuer die Tumortherapie', project Q1)

***Author contributions:*** GS, FS, KL and VM conceived and designed the research, GS, JZ and TS performed the simulations, GS analyzed the data, SS KL and VM provided the experimental data, GS FS and VM wrote the manuscript.

**Tables**

Table 1 Residue-specific PEG/water ratio

| Amino Acid | $PEGW_{AA}$ [bulk ratio] |
|---|---|
| S | 0.50 ± 0.17 |
| D | 0.66 ± 0.22 |
| E | 0.82 ± 0.11 |
| P | 0.82 ± 0.22 |
| W | 0.87 ± 0.26 |
| N | 0.89 ± 0.23 |
| K | 0.89 ± 0.15 |
| T | 0.98 ± 0.22 |
| A | 1.00 ± 0.23 |
| L | 1.00 ± 0.23 |
| G | 1.02 ± 0.22 |
| Q | 1.07 ± 0.21 |
| Y | 1.11 ± 0.19 |
| C | 1.15 ± 0.18 |
| H | 1.17 ± 0.23 |
| R | 1.29 ± 0.18 |
| F | 1.53 ± 0.23 |
| V | 1.55 ± 0.26 |
| I | 1.56 ± 0.23 |
| M | 2.11 ± 0.23 |

Figures

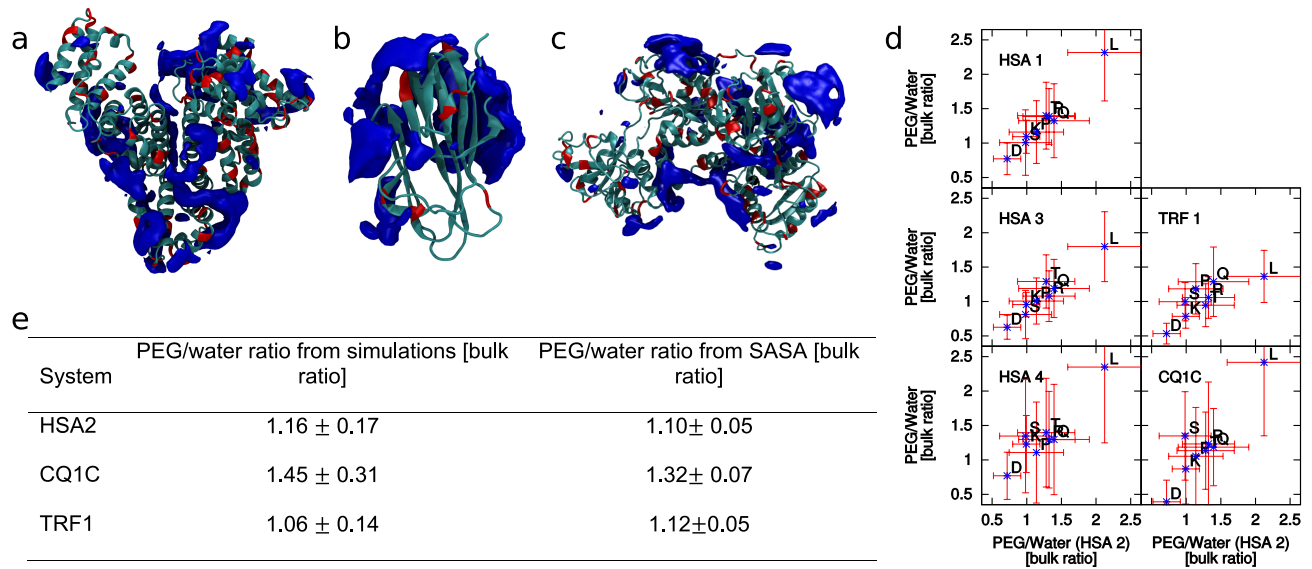

Figure 1 Cartoons of HSA (a), C1qC (b), TRF (c) highlighting the regions (dark blue) with an average PEG density twice as large as in the bulk. These regions are not uniformly distributed on the protein surface. They tend to be located away from negatively charged residues (red). (d) The residue-specific PEG/water ratios measured in several simulated systems and plotted versus those obtained for simulation HSA2 (see supplementary information) on the x axis to highlight the correlation. Only the amino acid types which contribute at least 3% of solvent accessible surface area in each of the simulated systems are considered. The correlation coefficients measured between each possible simulation pair are equal or larger than 0.7. The dependence of the PEG/water ratios on PEG concentration, density or box size (left column) is weak. The impact of the simulated protein on the PEG/water ratios (right column) is also limited. The error bars correspond to the standard deviations measured along the trajectories of each simulation. (e) PEG/water ratio in the 0.5 nm region around the protein averaged over the simulations (left column) and approximated using eq. (1)(right column).

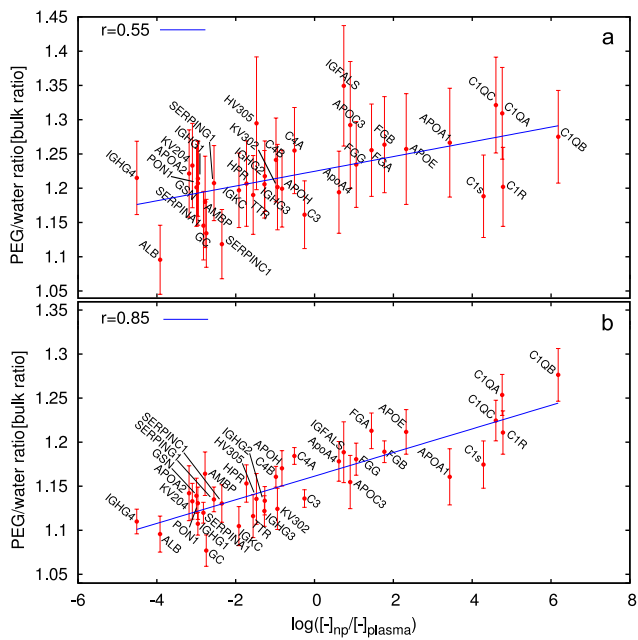

Figure 2 The PEG/water ratios (right hand side of eq. (5)) of the plasma proteins, estimated (a) using the $PEGW_{AA}$ from the simulations and (b) using the $PEGW_{AA}$ optimized through the bootstrap procedure (see supplementary information) are plotted versus the logarithm of the protein concentration ratio on the nanoparticle corona and in plasma. The blue line represents the best linear fit and the correlation coefficients (r) are reported in the upper left corner of the respective box. In (a) the error bars are obtained by propagating the error on the $PEGW_{AA}$. In (b) the error bars report the standard deviations measured over all the bootstrap set. Each protein is indicated with its gene name. See Table S2 for the corresponding full name.

*Supplementary information for*

# Protein corona composition of PEGylated nanoparticles correlates strongly with amino acid composition of protein surface

Giovanni Settanni, Jiajia Zhou, Tongchuan Suo, Susanne Schöttler, Katharina Landfester, Friederike Schmid, Volker Mailänder

### *Molecular dynamics simulations*

All the simulations were carried out using the program NAMD[1] and the charmm27 force field[2,3] with the extension for PEG[4]. Tip3p[5] was used as model for the explicit treatment of water. An integration time step of 1 fs was used across the simulations. Simulations were carried out using periodic boundary conditions. Pressure and temperature were maintained constant at 1atm and 300K, respectively, during the simulations using the Langevin piston algorithm and Langevin thermostat[6,7]. A cutoff of 1.2 nm was used for the non-bonded interactions with a switch function. Long range electrostatic interactions were treated using the smooth particle mesh Ewald (PME) method[8] with a grid spacing of about 0.1 nm. To prepare the PEG-water mixture, 64 PEG molecules (H-[O-CH$_2$-CH$_2$]$_n$-OH, with $n$ either 4 or 7) were placed on a 4x4x4 grid with 1. nm spacing between grid points. Then 1ns high temperature (700K) simulations in vacuum with damped electrostatics interactions (dielectric constant 200) were run to randomize the initial dihedral distribution of the PEG molecules. The PEG molecules were then immersed in a box of water molecules and sodium and chlorine ions were added to reach physiological concentration (0.15 M). Mixtures with different concentrations of PEG were obtained by changing the size of the water box surrounding the PEG molecules. The prepared mixtures were then equilibrated first at high temperature (373K) for 1.0 ns and then at 300K for 1.0 ns. The initial coordinates of the proteins were taken from the PDB (see Table S1 for the list of PDBids). Each protein was immersed in a box filled by replicating the coordinates of the PEG-water mixtures obtained before in the three space directions and removing mixture atoms in close contact with protein atoms. The final PEG concentration in the simulation boxes is reported in Table S1. The size of the boxes was large enough to leave at least 1.0 /nm from each protein atom and the box boundary. In the case of HSA, larger box sizes with 1.5 and 2.0 nm distances between protein and box boundary were simulated to investigate the dependence of the simulation results upon box size. The total charge of the systems was neutralized by changing an adequate number of water molecules into ions. The complete systems were then minimized using the steepest descent algorithm for 10000 steps with harmonic restraints on the heavy atoms of the proteins. Then the systems were equilibrated at room temperature and pressure for 1.0 ns during which the harmonic restraints were gradually removed and for 1.0 ns without restraints. Finally production runs were started with 4 or 5 replicas for each system. Most of the runs reached the 200ns time length (Table S1).

**Table S1 List of the performed simulations**

| System | SimName | PDBid | Box size (Å) | N. Atoms | PEG length | [PEG](g/ml) | Simulation time (ns) |
|---|---|---|---|---|---|---|---|
| HSA | HSA1 | 1A06 | 98.8 | 10088 | 4 | 0.08 | 4 x 200 |
|  | HSA2 |  | 98.2 | 99301 | 4 | 0.11 | 4 x 200 |
|  | HSA3 |  | 108.6 | 13413 | 4 | 0.12 | 4 x 200 |
|  | HSA4 |  | 109.0 | 13477 | 7 | 0.04 | 5 x 200 |
|  | HSA5 |  | 118.2 | 17254 | 4 | 0.12 | 5 x 100 |
| CQ1C |  | 2WNV | 77.0 | 47912 | 4 | 0.12 | 5 x 200 |
| Transferrin | TRF1 | 2HAV | 108 | 13200 | 4 | 0.11 | 5 x 200 |
|  | TRF2 |  | 108.7 | 13466 | 4 | 0.07 | 5 x 200 |
|  | TRF3 |  | 108.4 | 13330 | 7 | 0.07 | 5 x 200 |
|  | TRF4 |  | 108.8 | 13437 | 7 | 0.04 | 5 x 200 |

*Analysis of trajectories*

The direct PEG-protein interactions along the simulations were measured using the NAMD pair-interaction utility. The time series of these interactions reaches convergence during the course of the simulations with relaxation times between 10 and 40ns. The time series along the trajectories of the number of PEG and water heavy atoms found within 0.5 nm of each amino acid type was determined using the "pbwithin" selection command of VMD. The ratio of these numbers was compared to the ratio between all the PEG and water heavy atoms in the simulation box (bulk ratio). The autocorrelation function of this number relaxes in less than 10ns along all the analyzed simulations. Thus, the first 10ns of data were discarded and 10ns-long block averages were used for the determination of standard deviations. A PEG/water ratio larger than bulk for an amino acid implies the presence of an effective attractive interaction between PEG and the amino acid. On the other hand, amino acids with a PEG/Water ratio smaller than bulk exert and effective repulsion for PEG.

*Mass spectrometry*

The concentration of proteins in plasma and $PEG_{44}$-coated polystyrene (PS-PEG44) nanoparticles was obtained by mass spectrometry. This dataset has been published elsewhere (see Table S9 of supporting information of reference[10]) along with the experimental methods for mass spectrometry. In cases where the measurement in plasma was below the detectability threshold, the plasma concentration of the protein was taken from the literature[11] and converted to the same units as in the mass spectrometry data, taking the concentration of serum albumin as reference.

*Structural analysis*

Whenever they were available, the structures of the proteins with a non-zero concentration on the nanoparticles were collected from the PDB or from the database of homology models[12] when the homology with the template was larger than 40%. In cases where the protein is a subunit of a complex, only the structure of the corresponding subunit was retained. Overall 36 protein structures satisfied the criteria and were retained for analysis (Table S2). The solvent exposed surface area of each amino-acid of the collected proteins was measured using VMD and 0.14 nm as probe radius.

**Table S2 List of proteins included in the fitting data set**

| Protein | Gene/Name | PDBid+chain |
|---|---|---|
| Alpha-1-antitrypsin | SERPINA1 | 3NDD A |
| Antithrombin-III | SERPINC1 | 1E03 L |
| Apolipoprotein A-I | APOA1 | 3K2S A |
| Apolipoprotein A-II | APOA2 | 2OU1 |
| Apolipoprotein A-IV | APOA4 | 3S84 A |
| Apolipoprotein C-III | APOC3 | 2JQ3 |
| Apolipoprotein E | APOE | 2L7B |
| Beta-2-glycoprotein 1 | APOH | 1QUB |
| Complement C1q subcomponent subunit A | C1QA | 2JG8 A |
| Complement C1q subcomponent subunit B | C1QB | 2JG8 B |
| Complement C1q subcomponent subunit C | C1QC | 2WNV C |
| Complement C1r subcomponent | C1R | 1GPZ A |
| Complement C1s subcomponent | C1S | 4J1Y A |
| Complement C3 | C3 | 2A73 |
| Complement C4-A | C4A | 4FXG alpha |
| Complement C4-B | C4B | 4FXG beta |
| Fibrinogen alpha chain | FGA | 3GHG alpha |
| Fibrinogen beta chain | FGB | 3GHG beta |
| Fibrinogen gamma chain | FGG | 3GHG gamma |
| Gelsolin | GSN | 3FFN A |
| Haptoglobin-related protein | HPR | P00739 |
| Ig gamma-1 chain C region | IGHG1 | 1HZH H |
| Ig gamma-2 chain C region | IGHG2 | P01859 |
| Ig gamma-3 chain C region | IGHG3 | P01860 |
| Ig gamma-4 chain C region | IGHG4 | P01861 |
| Ig heavy chain V-III region BRO | HV305 | P01766 |
| Ig kappa chain C region | IGKC | 4XMP L |
| Ig kappa chain V-III region SIE | KV302 | P01620 |
| Ig kappa chain V-II region TEW | KV204 | P01617 |
| Insulin-like growth factor-binding protein complex acid labile subunit | IGFALS | P35858 |
| Plasma protease C1 inhibitor | SERPING1 | 2OAY |
| Protein AMBP | AMBP | 4ES7 A |
| Serum albumin | ALB | 1AO6 A |
| Serum paraoxonase/arylesterase 1 | PON1 | P27169 |
| Transthyretin | TTR | 4PVL |
| Vitamin D-binding protein | GC | 1KW2 A |

### *Fit of model via bootstrap procedure*

In eq. (5) the left hand side is obtained from mass spectrometry experiments and, in the right hand side, the $\text{FSASA}_{\text{AA}}(P)$ are obtained from the structural analysis described above. The $\text{PEGW}_{\text{AA}}$ are considered as the free parameters of the fit. The fit was carried out using a bootstrap approach, which allows for estimating the robustness of the results to changes in the fitted data set. The approach is as follows: the initial data set of 36 proteins was resampled 100 times, i.e. 100 new sets of 36 proteins were generated by randomly picking any protein in the original set, allowing for duplicates. The resampled sets contained from a minimum of 20 to a maximum of 30 unique proteins of the original set (median 25.5). For each resampled data set an optimal set of parameters $\text{PEGW}_{\text{AA}}$ was obtained by a simple null-temperature Monte Carlo procedure where the parameters were randomly modified and the new parameters were accepted only if they led to an increase in the correlation coefficient between the left and the right hand side of eq.(5). The correlation coefficient between two sets of values $\{X_i\}$ and $\{Y_i\}$ was defined as:

$$r = \frac{\sum^i X_i Y_i - \langle X \rangle \langle Y \rangle}{\sqrt{(\langle X^2 \rangle - \langle X \rangle^2)(\langle Y^2 \rangle - \langle Y \rangle^2)}}$$

A rescaling of the $\text{PEGW}_{\text{AA}}$ does not change the correlation coefficient between r.h.s. and l.h.s. of eq.(5). Thus, we rescaled the $\text{PEGW}_{\text{AA}}$ obtained for each resampled set, so that the average $\text{PEGW}_{\text{tot}}$ over the original set of proteins (not the resampled) equals 1. The extracted parameters are then rescaled so that the PEG/water ratio for ALB is the same as the one measured with the simulations. Finally, we computed the average over the resampled sets of the $\text{PEGW}_{\text{AA}}$ and their standard deviation (Table 1). Similarly we computed the average and the standard deviation over the resampled sets of the r.h.s. of eq.(5) (that is the $\text{PEGW}_{\text{tot}}(P)$) of each of the proteins in the original dataset, and plotted them versus the l.h.s. of eq. (5) (Figure 2). The correlation coefficient measured between the two data sets is 0.85. Extracting the optimal parameters for the full set of proteins improves the correlation coefficient to 0.87 but the obtained parameters are within a standard deviation from the average provided in Table 1.

## *Supplementary Figures*

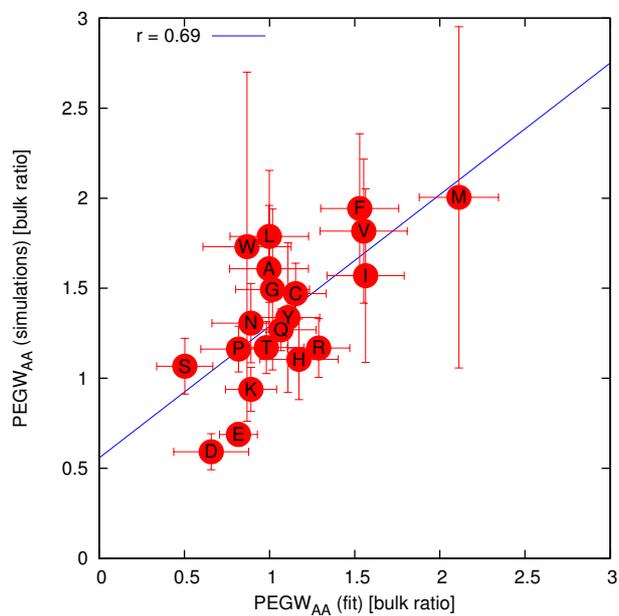

Figure S 1 The residue specific affinities for PEG ($PEGW_{AA}$) obtained from the fit are plotted vs the values obtained from the simulations. The correlation coefficient is reported in the top left corner of the figure. The blue line is the best linear fit.

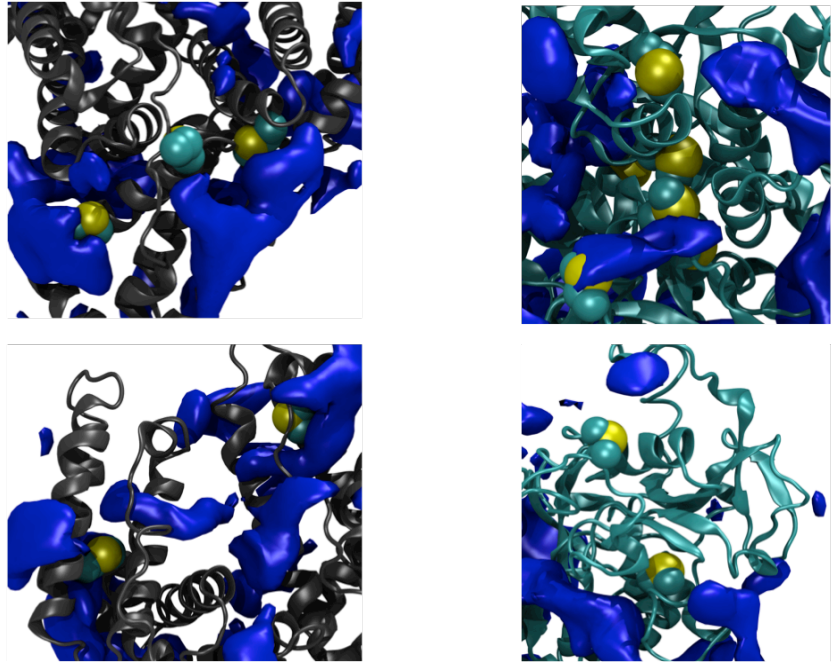

Figure S 2 Cartoons highlighting the presence of high density PEG regions (dark blue) around methionine residues (cyan and yellow balls) in HSA (left) and TRF (right)